\documentstyle[12pt]{article}
\def\be{\begin{equation}}
\def\ee{\end{equation}}
\def\ba{\begin{array}}
\def\ea{\end{array}}

\begin{document}
\title{ Implications of Configuration Mixing in The Chiral Quark Model
With SU(3) and Axial U(1) Breakings for Nucleon Spin-Flavor Structure.}
\author{Harleen Dahiya  and Manmohan Gupta    \\
{\it Department of Physics, Panjab University,Chandigarh-160 014,
 India.}}   
 \maketitle

\begin{abstract}
The implications of Chiral Quark Model  with SU(3) and axial U(1) symmetry
breakings as well as configuration mixing generated by
one gluon exchange forces ($\chi$QM$_{gcm}$) are discussed
in the context of proton flavor and spin structure  as
well as the hyperon $\beta$-decay data. 
Apart from reproducing the success of $\chi$QM with symmetry breaking,
it is  able to improve upon the agreement with data in
several cases such as,
$G_A/G_V, ~\Delta_8, ~<A_1^p>$ dependent on spin polarization
functions and ($\frac{2 \bar s}{u+d}$),
($\frac{2 \bar s}{\bar u+\bar d}$) and $f_s$ involving the quark
distribution functions.
\end{abstract}

It is well known that the chiral quark model ($\chi$QM)
{\cite{{manohar},{wein},{cheng}}}
with SU(3) symmetry is not only able to give a fair explanation of
``proton spin crisis'' {\cite{EMC}} but is also able to account for
the $\bar u-\bar d$ asymmetry {\cite{{GSR},{GSR1}}} as well as the
existence of significant strange
quark content $\bar s$ in the nucleon when the asymmetric
octet singlet couplings are taken into account  {\cite{st q}}.
Further, $\chi$QM with SU(3) symmetry is
also able to provide fairly satisfactory explanation for various
quark flavor contributions to the proton spin {\cite{eichten}}, baryon
magnetic moments {\cite{{cheng},{eichten}}} as well as the absence of
 polarizations of the
antiquark sea  in the nucleon {\cite{{song},{antiquark}}} .
However, in the case of hyperon decay parameters
 the predictions of the
$\chi$QM are not in tune with the data   {\cite{decays}}, for example,
in comparison to the experimental numbers   .21 and 2.17
the $\chi$QM with SU(3) symmetry predicts   $f_3/f_8$ and
$\Delta_3/\Delta_8$ to be $\frac{1}{3}$ and $\frac{5}{3}$ respectively.
It has been shown {\cite{{song},{cheng1}}} that when SU(3) breaking
 effects are taken into consideration within $\chi$QM, the
 predictions of the $\chi$QM regarding the above mentioned ratios
 have much better overlap with the data.

Recently it has been shown {\cite{hd}}  that the one gluon mediated
configuration mixing, within the premises of constituent quark model,
not only explains the
neutron charge radius squared ($<r_n^2>$) but is also able to
improve $G_A/G_V$ fit in comparison
with calculations without configuration mixing.
Therefore, it becomes interesting to examine,
within the $\chi$QM, the implications
of one gluon mediated configuration mixing for flavor and spin
structure of nucleon. 
 In particular we would like to examine
 the nucleon spin
polarizations and various hyperon $\beta$-decay constants,
 violation of Gottfried sum rule, strange
quark content in the nucleon, fractions of quark flavor etc.
in the $\chi$QM with configuration mixing and with
and without symmetry breaking.
Further, it would be intersting to examine whether a unified
fit could be effected for spin polarization functions as well as
quark ditribution functions or not.

The details of the $\chi$QM$_{gcm}$ 
have been discussed in the reference {\cite{hd}},
however for the sake of readability of manuscript, we summarize
the essential features of $\chi$QM$_{gcm}$.  
The basic process, in the $\chi$QM, is the
emission of a Goldstone Boson (GB) which further splits into $q \bar q$
pair, for example,

\be
  q_{\pm} \rightarrow GB^{0}
  + q^{'}_{\mp} \rightarrow  (q \bar q^{'})
  +q_{\mp}^{'}.
\ee
The effective Lagrangian describing interaction between quarks
and the octet GB and singlet $\eta^{'}$ is

\be
{\cal L} = g_8 \bar q \phi q,
\ee
where $g_8$ is the coupling constant,
\[ q =\left( \ba{c} u \\ d \\ s \ea \right)\]
and
\[ \phi = \left( \ba{ccc} \frac{\pi^0}{\sqrt 2}
+\beta\frac{\eta}{\sqrt 6}+\zeta\frac{\eta^{'}}{\sqrt 3} & \pi^+
  & \alpha K^+   \\
\pi^- & -\frac{\pi^0}{\sqrt 2} +\beta \frac{\eta}{\sqrt 6}
+\zeta\frac{\eta^{'}}{\sqrt 3}  &  \alpha K^0  \\
 \alpha K^-  &  \alpha \bar{K}^0  &  -\beta \frac{2\eta}{\sqrt 6}
 +\zeta\frac{\eta^{'}}{\sqrt 3} \ea \right). \]

SU(3) symmetry breaking is introduced by considering
different quark masses $m_s > m_{u,d}$ as well as by considering
the masses of  non-degenerate Goldstone Bosons
 $M_{K,\eta} > M_{\pi}$, whereas 
  the axial U(1) breaking is introduced by $M_{\eta^{'}} > M_{K,\eta}$
{\cite{{cheng},{song},{cheng1},{johan}}}.
The parameter a(=$|g_8|^2$) denotes the transition probability
of chiral fluctuation
or the splittings  $u(d) \rightarrow d(u) + \pi^{+(-)}$, whereas 
$\alpha^2 a$ denotes the probability of transition 
$u(d) \rightarrow s  + K^{-(0)}$.
Similarly $\beta^2 a$ and $\zeta^2 a$ denote the probability of
$u(d,s) \rightarrow u(d,s) + \eta$ and
$u(d,s) \rightarrow u(d,s) + \eta^{'}$ respectively.

The one gluon exchange forces {\cite{DGG}}
generate the mixing of the octet in $(56,0^+)_{N=0}$ with
the corresponding octets in $(56,0^+)_{N=2}$,
$(70,0^+)_{N=2}$ and  $(70,2^+)_{N=2}$
harmonic oscillator bands {\cite{Isgur1}}. The
corresponding wave function of the nucleon is given by
 
 \[|B>=(|56,0^+>_{N=0} cos \theta +|56,0^+>_{N=2} sin \theta)
 cos \phi \] 
\be
 +(|70,0^+>_{N=2} cos \theta +|70,2^+>_{N=2}  sin \theta)
 sin \phi.
 \ee
In the above equation it should be noted that
$(56,0^+)_{N=2}$ does not effect the
spin-isospin structure of  $(56,0^+)_{N=0}$, therefore the mixed
nucleon wave function can be expressed in terms of  $(56,0^+)_{N=0}$
and  $(70,0^+)_{N=2}$, which we term as non trivial mixing
{\cite{mgupta1}} and is given as

\begin{equation}
\left|8,{\frac{1}{2}}^+ \right> = cos \phi |56,0^+>_{N=0}
+ sin \phi|70,0^+>_{N=2},
\end{equation}
 where
 \be
 |56,0^+>_{N=0,2} = \frac{1}{\sqrt 2}(\chi^{'} \phi^{'} +
\chi^{''} \phi^{''}) \psi^{s},
\ee

\be
|70,0^+>_{N=2} =  \frac{1}{2}[(\psi^{''} \chi^{'} +\psi^{'}
\chi^{''})\phi^{'} + (\psi^{'} \chi^{'} -\psi^{''} \chi^{''})
\phi^{''}].
\ee
 The spin and isospin wave functions, $\chi$ and $\phi$, are
given below
    \[\chi^{'} =  \frac{1}{\sqrt 2}(\uparrow \downarrow \uparrow
    -\downarrow \uparrow \uparrow),~~~  \chi^{''}
    =  \frac{1}{\sqrt 6} (2\uparrow \uparrow \downarrow
  -\uparrow \downarrow \uparrow
  -\downarrow \uparrow \uparrow), \] \\
\[\phi^{'}_p = \frac{1}{\sqrt 2}(udu-duu),~~~
\phi^{''}_p = \frac{1}{\sqrt 6}(2uud-udu-duu),\]
\[\phi^{'}_n = \frac{1}{\sqrt 2}(udd-dud),~~~
 \phi^{''}_n = \frac{1}{\sqrt 6}(udd+dud-2ddu).\]
 For the definition of the spatial part of the wave function,
 ($\psi^{s}, \psi^{'}, \psi^{''})$ as well as the
 definitions of the spatial overlap integrals we  refer the
 reader to reference {\cite{{mgupta1},{yaouanc}}.

 The contribution to the proton spin defined through the equation
 \be
\Delta q =q_\uparrow - q_\downarrow  +\bar{q}_\uparrow
 - \bar{q}_\downarrow,
 \ee
using Equation(4) and following Linde $et.at.$ {\cite{johan}}, can
be expressed as 

\be
   \Delta u ={cos}^2 \phi \left[\frac{4}{3}-\frac{a}{3}
   (7+4 \alpha^2+ \frac{4}{3} \beta^2
   + \frac{8}{3} \zeta^2)\right]
   + {sin}^2 \phi \left[\frac{2}{3}-\frac{a}{3} (5+2 \alpha^2+
  \frac{2}{3} \beta^2 + \frac{4}{3} \zeta^2)\right], 
\ee

\be
  \Delta d ={cos}^2 \phi \left[-\frac{1}{3}-\frac{a}{3} (2-\alpha^2-
  \frac{1}{3}\beta^2- \frac{2}{3} \zeta^2)\right]  + {sin}^2 \phi
  \left[\frac{1}{3}-\frac{a}{3} (4+\alpha^2+
  \frac{1}{3} \beta^2 + \frac{2}{3} \zeta^2)\right], 
\ee
and

\be
   \Delta s ={cos}^2 \phi \left[-a \alpha^2\right]+
 {sin}^2 \phi \left[-a \alpha^2 \right].
\ee

The  SU(3) symmetric calculations can easily be obtained from
Equations (8), (9), (10) by considering $\alpha, \beta = 1$. The
corresponding equations can be expressed as

 \be
   \Delta u ={cos}^2 \phi \left[\frac{4}{3}
   -\frac{a}{9} (37+8 \zeta^2)\right]  + {sin}^2 \phi
  \left[\frac{2}{3}-\frac{a}{9} (23+4 \zeta^2)\right], 
\ee

\be
  \Delta d ={cos}^2 \phi \left[-\frac{1}{3}
  -\frac{2a}{9} (\zeta^2-1)\right]  + {sin}^2 \phi
  \left[\frac{1}{3}-\frac{a}{9} (16+2 \zeta^2)\right], 
\ee
and

\be
   \Delta s = -a.
\ee

After having examined the effect of one gluon exchange
inspired configuration mixing on the spin polarizations of various quarks
$\Delta u,~ \Delta d$ and $\Delta s$, we can calculate the
following quantities

\be
G_A/G_V = \Delta_3=\Delta u-\Delta d,
\ee

\be
\Delta_8=\Delta u+\Delta d-2 \Delta s,
\ee
The asymmetry $<A_1^p>$
measured in deep inelastic scattering 

\be
\left <A_1^p \right >=2<x> \frac{\sum {e_q^2 \Delta q}}
{\sum {e_q^2 q}}.
\ee
can also be calculated in terms of the spin polarization functions.

Similarly the hyperon $\beta$ decay data
{\cite{{PDG},{close},{hyp1},{hyp2}}} can
also be expressed in terms of the spin polarization functions, for example,

\be
\Delta_3=\Delta u-\Delta d=F+D,
\ee

\be
\Delta_8=\Delta u+\Delta d-2 \Delta s=3F-D.
\ee

Before we present our results it is perhaps desirable to discuss
certain aspects of the symmetry breaking parameters employed here.
As has been considered by Cheng and Li {\cite{cheng}}, the singlet
octet symmetry breaking parameter
$\zeta$ is related to $\bar u- \bar d$ asymmetry
{\cite{{GSR},{GSR1}}, we have also
taken $\zeta$ to be responsible for the $\bar u-\bar d$ asymmetry
in the $\chi$QM with SU(3) symmetry breaking and configuration
mixing. Further, we have used the relation
$\zeta=-0.7-\frac{\beta}{2}$ in order to keep a=0.1 as has
been used by various authors
{\cite{{cheng},{eichten},{song},{cheng1},{johan}}}.
With these restrictions on
parameters, we have carried out a $\chi^2$ minimization for spin
as well as quark distribution functions.

In Table 1 we have presented
the results of our calculations pertaining to spin polarization
functions $\Delta u,~ \Delta d,~ \Delta s$ whereas in Table 2 the
corresponding hyperon $\beta$-decay parameters dependent on the
spin distribution functions have been presented.
Since one of the purpose of the present calculation is to compare the
present results with those of the results corresponding to only SU(3)
breaking, therefore we have included in  Table 1 and 2  the results
of Song $et.al.$ {\cite{song}}. The corresponding
results of Cheng and Li {\cite{cheng1}} have not been included
because they have not fitted $\bar u/\bar d$ through the singlet octet
SU(3) breaking parameter $\zeta$.
In Table 1 and 2 we have also included the results of NRQM and $\chi$QM
with SU(3) symmetry and the corresponding results after including
configuration mixing. This has been carried out to understand the
implications of configuration mixing in effecting the fit.

\begin{table}
{\small
\begin{center}
\begin{tabular}{|c|c|c|c|c|c|c|c|}       \hline
Parameter &Expt  & NRQM  & NRQM$_{gcm}{\cite{hd}}$ &  $\chi$QM   &  $\chi$QM  &
$\chi$QM$_{gcm}$    &  $\chi$QM$_{gcm}$   \\ 
  & value  & results & & with SU(3) & with  SU(3)  & with SU(3)  &
  with SU(3)  \\
& & & & symmetry &  symmetry    & symmetry  & symmetry \\
& & & &    & breaking   &     & breaking \\ \cline{5-6}      \hline

 & & & & $\alpha=1$ &  $\alpha=.5$ &  $\alpha=1$  & $\alpha=.4$   \\ 
 & & & & $\beta=1${\cite{cheng}} &   $\beta=1$ {\cite{song}}  & $\beta=1$  &  $\beta=.7$  \\ 
 \hline

$\Delta u$ &  0.85  $\pm$ 0.05 {\cite{adams}} & 1.33 & 1.25 & .79 &
.86 &  .77 & .90, ~.86$^*$ \\

$\Delta d$  & -.41 $\pm$ 0.05 {\cite{adams}} & -0.33 & -0.26 & -0.32 &
-0.34 &  -0.27 & -0.32,~ -0.36$^*$  \\

$\Delta s$   & -0.07 $\pm$ 0.05 {\cite{adams}}  & 0 & 0 &  -0.10 &
-0.05 &  -0.10 & -0.02, ~-0.06$^*$  \\

$G_A/G_V$ & 1.26 $\pm$ .0028 {\cite{PDG1}} & 1.66 & 1.51 &
1.11 & 1.20 &  1.04 & 1.22, ~1.22$^*$   \\

$\Delta_8$& .58 $\pm$.025 {\cite{PDG}} & 1 & .99 &.67 &
.62 &  .70 &  .61, ~.62$^*$  \\          

$\left <A_1^p \right >$ & 0.40 $\pm$ 0.10 {\cite{e143}} & - & - & .28 &
.35 &  .31  & .38, ~.36$^*$  \\     \hline

\end{tabular}
\end{center}
* {\small Values after inclusion of the contribution from anomaly 
{\cite{anomaly}}.}
\caption{The calculated values  of
spin polarization functions $\Delta u, ~\Delta d, ~\Delta s$,
and quantities dependent  on these: $G_A/G_V, ~\Delta_8$
and  $<A_1^p>$. The last two columns respectively list the values in
the $\chi$QM$_{gcm}$ with SU(3) symmetry and with SU(3) $\times$ U(1)
symmetry breakings for the values of $\alpha$ and  $\beta$ obtained
by $\chi^2$ minimization.}}
\end{table}

\begin{table}
{\small
\begin{center}
\begin{tabular}{|c|c|c|c|c|c|c|c|}       \hline
Parameter &Expt  & NRQM  & NRQM$_{gcm}{\cite{hd}}$ &
$\chi$QM  &  $\chi$QM with  & $\chi$QM$_{gcm}$ & $\chi$QM$_{gcm}$   \\ 
  & value   & results&  & with SU(3)  &
 with  SU(3)  & with SU(3)  & with SU(3)  \\
& & & & symmetry &  symmetry    & symmetry & symmetry \\
& & & &    & breaking    &     & breaking \\ \cline{5-6}

 \hline
 & & & & $\alpha=1$ &  $\alpha=.5$ &  $\alpha=1$  & $\alpha=.4$   \\ 
 & & & & $\beta=1$ {\cite{cheng}}&   $\beta=1$ {\cite{song}}  &
 $\beta=1$  &  $\beta=.7$  \\    \hline

F+D &  1.26 $\pm$ .0028 & 1.66 & 1.51 & 1.11 & 1.20 &  1.04 & 1.22   \\

F+D/3 & .718  $\pm$ .015 & 1 & .92 & .67 & .70 &  .64 & .71  \\

F-D & -.34  $\pm$ .017 & -0.33 & -0.26 & -0.22 & -0.29 & -0.17 & -0.30 \\

F-D/3 &  .25 $\pm$ .05 & .33 & .33 & .22 & .21 &  .23 & .21   \\

F/D &  .575  $\pm$ .016  & .67 & .71 & .67 & .61 & .72 & .60   \\

F &  .462 & .665 & .625 &.445 & .455 &  .435 & .46  \\

D &  .794 & 1 & .885 & .665 & .745 &  .605 & .76   \\   \hline
\end{tabular}
\end{center}
\caption{The last two columns respectively list the calculated values  
the hyperon $\beta$ decay data in the $\chi$QM$_{gcm}$ with SU(3) symmetry
and with SU(3) $\times$ U(1) symmetry breakings for the values of
$\alpha$ and  $\beta$ obtained by $\chi^2$ minimization.}}
\end{table}

A general look at Table 1  makes it clear that we have
been able to get an excellent fit to the spin polarization
data for the values of symmetry
breaking parameters $\alpha=.4,~ \beta=.7$ obtained by
$\chi^2$ minimization.
It is perhaps desirable to mention that the spin distribution functions
$\Delta u, ~\Delta d, ~\Delta s$ show much better agreement with data
when the contribution of anomaly {\cite{anomaly}} is included.
Similarly in Table 2 we find that the success of the fit obtained with
$\alpha=.4,~ \beta=.7$ hardly leaves anything to desire. The agreement is
striking in the case of parameters F+D/3 and F. We therefore conclude
that the $\chi$QM with SU(3) and axial  U(1) symmetry breakings along with
configuration mixing generated by one gluon exchange forces provides
a satisfactory description of the spin polarization functions and the
hyperon decay data.

 In order to appreciate the role of
configuration mixing in effecting the fit, we first  compare the
results of NRQM with those of NRQM$_{gcm}$ {\cite{hd}}.
One observes that
configuration mixing corrects the result of the quantities
in the right direction but this is not to the desirable level.
Further, in order to understand the role of configuration mixing
and SU(3) symmetry with and without breaking  in $\chi$QM,
we can compare the results of $\chi$QM with SU(3)
symmetry to those of $\chi$QM$_{gcm}$ with SU(3)
symmetry. Curiously $\chi$QM$_{gcm}$ compares unfavourably with  $\chi$QM
in case of most of the calculated quantities.
This indicates that configuration mixing alone is not enough to generate
 an appropriate fit in  $\chi$QM.
However when  $\chi$QM$_{gcm}$ is used with SU(3) and axial U(1) symmetry 
breakings
then the results show uniform improvement over the corresponding
results of $\chi$QM with  SU(3) and axial U(1) symmetry breakings \cite{song}.
In particular there is discernible improvement in the case of
$G_A/G_V, ~\Delta_8$ and $<A_1^p>$.
To summarize the  discussion of these results,
one finds that both configuration mixing
and symmetry breaking are very much needed to fit the data within
$\chi$QM.
It should also be borne in
mind that the improvement in the above mentioned case is while
maintaining the success of  $\chi$QM with SU(3) symmetry breaking.


In view   of the fact that flavor structure of nucleon is not affected
by configuration mixing, therefore it would seem that the results of  $\chi$QM
with SU(3) breaking will be exactly similar to those of  $\chi$QM$_{gcm}$
with SU(3) breaking. However, as mentioned earlier, one of the purpose of
the present communication is to have a unified fit to spin polarization
functions as well as quark distribution functions, therefore it is
interesting to compare our results with those obtained in $\chi$QM
with and without symmetry breaking.
To that end, we first mention the quantities which we
have calculated.The basic quantities of interest in this case are
the unpolarized quark distribution functions,
particularly the antiquark contents given as under {\cite{johan}}

\be
\bar u =\frac{1}{12}[(2 \zeta+\beta+1)^2 +20] a,
\ee

\be
\bar d =\frac{1}{12}[(2 \zeta+ \beta -1)^2 +32] a,
\ee

\be
\bar s =\frac{1}{3}[(\zeta -\beta)^2 +9 {\alpha}^{2}] a,
\ee
For the quark number in the proton, we have
\be
u=2+\bar u, ~~~d =1+\bar d, ~~~ s=\bar s.
\ee
There are important experimentally
measurable quantities dependent on the above distributions.
The deviation from the Gottfried sum rule {\cite{GSR1}} is one such
quantity which measures the 
 asymmetry between the $\bar u$ and $\bar d$ quarks in
the nucleon sea. In the $\chi$QM the deviation of
Gottfried sum rule from 1/3rd is expressed as

\be
\left [\int_{0}^{1} dx \frac{F_2^p(x)-F_2^n(x)}{x} -\frac{1}{3}
\right ] =\frac{2}{3} (\bar u-\bar d).
\ee
Similarly the $\bar u/\bar d$ which can be measured through the
ratio of muon pair production cross sections $\sigma_{pp}$  
and $\sigma_{pn}$ is also an important   parameter which gives
an insight into the $\bar u, ~\bar d$ content {\cite{baldit}}.
The other quantities of interest is the quark flavor fraction in a 
proton, $f_q$,  defined as

\be
f_q=\frac{q+\bar q}{[\sum_{q} (q+\bar q)]}
\ee
where q's stand for the quark numbers in the proton.
Also we have calculated the ratio of the total strange sea to
the light antiquark contents given by

\be
\frac{2 \bar s}{\bar u+ \bar d}
\ee
and the ratio of the total strange sea to the light quark
contents given by

\be
\frac{2 \bar s}{u+d}.
\ee

The above mentioned quantities based on quark distribution functions
have been calculated using the set of parameters,
$\alpha=.4$ and $\beta=.7$, which minimizes the $\chi^2$ fit for the spin
distribution functions and quark distribution functions.
The results of our calculations are presented in Table 3.
The general survey of Table 3 immediately makes it clear that the 
success achieved in the case of spin polarization functions is
very well maintained in this case also. Apparently it would seem
that $\chi$QM$_{gcm}$ with SU(3) symmetry breaking would not
add anything to the success in
$\chi$QM with SU(3) symmetry breaking. However, a closer look
at the table indicates that  we have been able to improve upon
the results of reference {\cite{song}} without any further input.
We would again like to mention here that we have not included
the results of Cheng $et.al.$ {\cite{cheng1}} for comparison because
they have not fitted the value of $\bar u/\bar d$ for the value
of the singlet octet breaking parameter $\zeta$.
In almost all the quantities, which have been measured experimentally,
our fit leaves hardly anything to be desired,
in contrast to the results of  {\cite{song}}. In comparison with
the results of {\cite{song}} our results show considerable
improvement in the case of ratio of the total strange sea
to the light antiquark contents ($\frac{2 \bar s}{\bar u+\bar d}$)
whereas there is a big improvement in the case of 
 ratio of the total strange sea to the light  quark contents
($\frac{2 \bar s}{u+d}$) and the strange flavor fraction ($f_s$).

\begin{table}
{\small
\begin{center}
\begin{tabular}{|c|c|c|c|c|c|c|}       \hline
Parameter &Expt  & NRQM  &
$\chi$QM   &
  $\chi$QM    &
$\chi$QM$_{gcm}$     \\ 
  & value  & results & with SU(3)  &
  with SU(3)  & with SU(3)    \\
& & & symmetry & symmetry   & symmetry \\
& & &    & breaking   
  & breaking \\ \cline{5-6}

 \hline

 & & &$\alpha=1$ &  $\alpha=.5$ &  $\alpha=.4$   \\ 
 & & & $\beta=1$ {\cite{cheng}} &   $\beta=1$ {\cite{song}}&
   $\beta=.7$  \\  \hline

$\bar u$ &  & & .168 &  .168 &  .168  \\

$\bar d$ &  & & .315 &  .315 &  .315  \\

$\bar s$ &  & & .46 & .34 &   .15  \\

$\bar u-\bar d$ &-.147 $\pm$ .024 {\cite{GSR}} & 0 &  -.147  &
-.147 & -.147  \\

$\bar u/\bar d$ & 0.51 $\pm$ 0.09 {\cite{baldit}} & 1 & .53 &
.53 &  .53  \\
 
$I_G$ & .235  $\pm$ .005 & 0.33 & .235 & .235 & .235   \\ 

$\frac{2 \bar s}{\bar u+ \bar d}$ & .477 $\pm$ .051 {\cite{ao}} &
&1.9 &
 1.4 &  .62  \\

$\frac{2 \bar s}{u+d}$ & .099 $\pm$ .009 {\cite{ao}} & 0 & .26 &
 .20 &   .09  \\

$f_u$ &  & & .48 & .51 &   .55  \\

$f_d$ & & & .33 & .35 &  .38  \\

$f_s$ &  .10 $\pm$ 0.06 {\cite{grasser}} & 0 & .19 &
.15 &   .07  \\

$f_3=$  & & & .15 & .16 &   .17  \\
$f_u-f_d$  &  & & & &  \\

$f_8=$ & & & .43 & .54 &    .79  \\
$f_u+f_d-2 f_s$ &   & & & &   \\

$f_3/f_8$ & .21 $\pm$ 0.05 {\cite{cheng}} & .33 & .33 &
.23 &   .21 \\
\hline

\end{tabular}
\end{center}
\caption{The last column lists the quark distribution
functions and  other dependent quantities as calculated
in the $\chi$QM$_{gcm}$ with SU(3) symmetry breaking.}} 
\end{table}

In conclusion we would like to mention that the
$\chi$QM with configuration mixing generated by one gluon mediated
forces ($\chi$QM$_{gcm}$) {\cite{hd}}
with SU(3) and axial U(1) symmetry breakings, is not only able to
fit the data regarding
spin polarization functions and the quark
distribution functions but is also able to improve upon the results of 
 $\chi$QM with SU(3) symmetry breaking. In particular
it is able to give an improved fit in the case of
$G_A/G_V, ~\Delta_8$ and $<A_1^p>$
as well as for     $\frac{2 \bar s}{u+d}$,
$\frac{2 \bar s}{\bar u+\bar d}$ 
and $f_s$, in contrast to the $\chi$QM with SU(3) symmetry
breaking  {\cite{song}}.
  
\vskip .2cm
  {\bf ACKNOWLEDGMENTS}\\
H.D. would like to thank CSIR, Govt. of India, for
 financial support and the chairman,
 Department of Physics, for providing facilities to work
 in the department.

\end{document}